\documentstyle[12pt]{article}

\begin{document}
\thispagestyle{empty}

\begin{center}
               RUSSIAN GRAVITATIONAL SOCIETY\\
               INSTITUTE OF METROLOGICAL SERVICE \\
               CENTER OF GRAVITATION AND FUNDAMENTAL METROLOGY\\

\end{center}
\vskip 4ex
\begin{flushright}
                                         RGS-VNIIMS-010/97
                                         \\ gr-qc/9710120

 \end{flushright}
\vskip 15mm

\begin{center}
{\large\bf CAUSALITY IN TOPOLOGICALLY NONTRIVIAL 
SPACE-TIMES\footnote{Submitted to the Proceedings of MG8 Conference} }

\vskip 5mm
{\bf
M.Yu. Konstantinov }\\
\vskip 5mm
     {\em VNIIMS, 3-1 M. Ulyanovoy str., Moscow, 117313, Russia}\\
     e-mail:  \\
\end{center}
\vskip 10mm

\begin{abstract}
The problems causality and causality violation in topologically nontrivial
space-time models are considered. To this end the mixed boundary problem
for traversable wormhole models is formulated and the influence of the
boundary conditions on the causal properties of space-time is analyzed
\end{abstract}

\vskip 10mm


\vskip 50mm

\centerline{Moscow 1997}

\vskip 20mm

\pagebreak


The problems of causality and causality violation in topologically
nontrivial space-time models are considered in connection with (1) numerous
declarations about ''absurdly easy wormhole transformation into the time
machine''~\cite{morris,idn,frolnov,visser}, which contradict to well-known
theorems about global hyperbolicity and Cauchy problem~\cite{hawk,fisher},
and (2) statements about existence of physical ''paradoxes'' in the presence
of time machine~\cite{krasnikov} as well as the statements about necessity
of introduction for such models an additional ''self-consistency
conditions'' to avoid these paradoxes~\cite{idn}.

To describe space-time model with traversable wormhole it is necessary to
use minimum two maps: one (or more) map for description of the wormhole
interior and one (or more) map for external space-time. Consider the
simplest case when the space section of the wormhole has topology of the
direct product $S^2\times I$, where $I$ is a closed interval $-L_1<\xi ^1<L_2
$ and $S^2$ is a two sphere with coordinates $(\xi ^2,\ \xi ^3)$. Let
external space-time is described by coordinates $\{t,\ x^i\}$ and has metric
\begin{equation}
\label{eq:extmetr}ds_{ext}^2=dt^2-\gamma _{ij}dx^idx^j,
\end{equation}
and interior space-time, which is described by coordinates $\{\tau ,\ \xi
^j\}$, has metric
\begin{equation}
\label{eq:intmetric}ds_{int}^2=a^2(\tau ,\xi )d\tau ^2-2b_i(\tau ,\xi )d\tau
d\xi ^i-\widetilde{\gamma }_{ij}(\tau ,\xi )d\xi ^id\xi _j,
\end{equation}
where $-\infty <t,\ \tau <\infty $, and $a^2(\tau ,\xi )>0$ because of the
supposition about traversability of the wormhole in both direction. In
general case the correspondence between internal and external coordinates
near the mouths of the wormhole are given by the following equations
\begin{equation}
\label{eq:ext}t_{left}=\tau ,\ \ x_{left}^i=x_{left}^i(\xi ^1,\xi ^2,\xi
^3),\ \ \ \ t_{right}=t_r(\tau ),\ \ x_{right}^i=x_{right}^i(\tau ,\xi
^1,\xi ^2,\xi ^3)
\end{equation}
Equations~(\ref{eq:ext}) define the topology of space-time.
They show in particular that for fixed $\tau $ wormhole connects the
points of external space-like hypersurface $(t_1(\tau ),M_{ext1}^3)$ with
the points of external space-like hypersurface $(t_2(\tau ),M_{ext2}^3)$,
where $ t_1\neq t_2$ in general case. These equations
induce the following boundary conditions for the components of internal
metric of the wormhole:
\begin{equation}
\label{eq:asympt1}a^2(\tau ,\xi )=%
\cases{1 & for $\xi ^1 \rightarrow -L_1$; \cr
\alpha ^2\cdot (1-v_iv^i) & for $\xi ^1 \rightarrow L_2$, \cr
}
\end{equation}

\begin{equation}
\label{eq:asympt2}\beta _i=%
\cases{
0 & for $\xi ^1 \rightarrow -L_1$ \cr
\gamma _{kl}\frac{\partial x_{right}^k}
{\partial \tau }\frac{\partial x_{right}^l}{\partial
\xi ^i} & for $\xi ^1 \rightarrow L_2$, \cr
}
\end{equation}
and
\begin{equation}
\label{eq:asympt3}\widetilde{\gamma }_{ij}=%
\cases{
\gamma _{kl}\frac{\partial x_{left}^k}
{\partial \xi ^i}\frac{\partial x_{left}^l}{\partial
\xi ^j} & for $\xi ^1 \rightarrow -L_1$, \cr
\gamma _{kl}\frac{\partial x_{right}^k}
{\partial \xi ^i}\frac{\partial x_{right}^l}{\partial
\xi ^j} & for $\xi ^1 \rightarrow L_2$, \cr
}
\end{equation}
where $\alpha =dt_r/d\tau $, and $v_l=\frac 1\alpha \gamma _{kl}\frac{%
\partial x_{right}^k}{\partial \tau }$.

Equations~(\ref{eq:ext})-(\ref{eq:asympt3}) together with condition $%
a^2(\tau ,\xi )>0$, which follows from the supposition about traversability
of the wormhole in both direction, and standard initial conditions for
internal and external metric, which near the left mouth must also satisfy to
the conditions analogous to Eqs.~(\ref{eq:asympt1})-(\ref{eq:asympt3}), form
the mixed boundary problem for Einstein equations. The mixed boundary
problems for the source fields are formulated by analogous manner. In
general case these boundary problems cannot be reduced to Cauchy problem.
So, the models with causality violation have non-evolutionary nature. In
particular, in opposite to the statements~\cite{morris,idn,visser} these
models can not be considered as ''transformation'' of some initial
configuration into the time machine.

It is easy to see that equations~(\ref{eq:asympt1})-~(\ref{eq:asympt3}) and
analogous conditions for other fields are the part of definition of the
corresponding geometrical objects on manifold. In particular, they provide
the self-consistency of solutions and the absence of any "paradoxes" in
space-time models with causality violation.

On the other hand because of the non-evolutionary nature of space-time
models with causality violation, which solve the general mixed boundary
problem are formulated above, the question about the physical sense of
models with causality violation is reduced to the question about the
physical sense of boundary conditions~(\ref{eq:ext})-(\ref{eq:asympt3}),
which must be given in general case in the points are separated bi time-like
intervals. The existence of non-singular solutions of the above mixed
boundary problem for space-time models with non-trivial causal structure is
also unresolved problem.

It is clear that the causal properties of space-time with traversable
wormhole are defined in general case both by the boundary conditions~(\ref
{eq:ext}) and by the field equations. Nevertheless at least in two
particular cases the causal structure of space-time is independent both on
the motion of the mouths of the wormhole and on other physical processes.

Namely, it is easy to see that if $t_{left}=t_{right}=\tau $ in equations~(%
\ref{eq:ext}) then the causality violation is impossible independenty on the
wormhole' mouths motion in the exterior space or on the other physical
processes in space-time. Moreover, if the length of the wormhole's handle is
short enough then in the external space-time (1) the absolute
synchronization of events near the mouths of the wormhole as well as the
preferable class of reference frames is exist; (2) the motion of the
wormhole' mouths is absolute in difference with the motion of the material
bodies. In particular, the same motion of the mouths as the bodies motion in
well known ''twins paradox'' of special relativity does not lead to
causality violation~\cite{myuk92,myuk95} in opposite to conclusions of~\cite
{morris,idn}.

Another particular case corresponds to the spherical wormhole with immovable
mouths (i.e. if $x_{right}^i=x_{right}^i(\xi ^1,\xi ^2,\xi ^3)$ in
equations~(\ref{eq:ext}) ) which is connected to external Minkowski
space-time. Direct calculations show that in this case both the structure
and properties of energy-momentum tensor in the wormhole's interior for the
models with causality violations are the same as in non-causal case which
were considered in~\cite{thorne}. Hence the causal structure of such model
is defined completely by conditions~(\ref{eq:ext}).

Analysis of the other kinds of space-time models with causality violation
shows that they also satisfy to the boundary or mixed boundary problem which
is similar to the considered above and do not contain any paradoxes.

\section*{Acknowledgments}

This work was supported in part by the Russian Ministry of Science and the
Russian Fund of Basic Research (grant N 95-02-05785-a).

\end{document}